\documentclass[aps,prb,reprint,superscriptaddress, nolongbibliography]{revtex4-2}
\usepackage[dvipsnames]{xcolor}
\usepackage{graphicx}
\usepackage{amsmath,amssymb}
\usepackage{braket}
\usepackage{simpler-wick}
\usepackage{makerobust}
\usepackage{gensymb}
\usepackage{bm}

\definecolor{goodred}{rgb}{0.7,0,0}
\usepackage[colorlinks,urlcolor=blue,citecolor=blue,linkcolor=goodred]{hyperref}

\def\equationautorefname~#1\null{%
	Eq.\,(#1)\null
}

\def\figureautorefname~#1\null{%
	Fig.\,#1\null
}

\def\sectionautorefname~#1\null{%
	Sec.\,#1\null
}

\begin{document}

\newcommand{\nn}{\nonumber}
\newcommand{\ie}{\emph{i.e.}}
\newcommand{\eg}{\emph{e.g.}}

\newcommand{\CD}{\hat \nabla}
\newcommand{\curl}{\text{curl}}
\renewcommand{\vec}[1]{{\bm{#1}}}


\title{Josephson anomalous vortices}

\author{Dan Crawford}
\affiliation{Department of Physics and Nanoscience Center, University of Jyväskylä,
P.O. Box 35 (YFL), FI-40014 University of Jyväskylä, Finland}
\author{Stefan Ili\'{c}}
\affiliation{Department of Physics and Nanoscience Center, University of Jyväskylä,
P.O. Box 35 (YFL), FI-40014 University of Jyväskylä, Finland}
\author{Pauli Virtanen}
\affiliation{Department of Physics and Nanoscience Center, University of Jyväskylä,
P.O. Box 35 (YFL), FI-40014 University of Jyväskylä, Finland}
\author{Tero T. Heikkilä}
\affiliation{Department of Physics and Nanoscience Center, University of Jyväskylä,
P.O. Box 35 (YFL), FI-40014 University of Jyväskylä, Finland}

\date{\today}

\begin{abstract}
	We show that vortices with circulating current, related with odd-frequency
	triplet pairing, appear in Josephson junctions where the barrier is a weak
	ferromagnet with strong spin-orbit coupling. By  symmetry analysis  we show
	that there is an additional term --- a rotary invariant --- in the
	superconducting free energy which allows for magnetoelectric effects even
	when the previously considered Lifshitz invariant vanishes. Using a
	microscopic model based on a modified Usadel equation incorporating those
	effects, we show that the size, shape, and position of these vortices can be
	controlled by manipulating Rashba spin-orbit coupling in the weak link, via
	gates, and we suggest that these vortices could be detected via scanning
	magnetometry techniques. We also show that the transverse triplet components
	of the pairing amplitudes can form a texture.
\end{abstract}

\maketitle


{\it Introduction---} Vortices in superconductors have remained a vibrant area
of research since their discovery in the 1950s. In the original work by
Abrikosov \cite{abrikosov_magnetic_1957}, vortices appear in type II
superconductors threaded by a flux quantum as a result of the superconducting
phase winding by $2\pi$ around a singularity. This produces a signature
circulating current with a normal core. Superconducting heterostructures allow
for a rich variety of other kinds of vortices. For example, in Josephson
junctions coreless, elongated vortices can arise due to solitons in the
superconducting phase \cite{ustinov_solitons_1998}.
Vortices with cores can also appear in normal metals in proximity to a
superconductor \cite{stolyarov_expansion_2018}, and also in junctions due to
destructive interference of the pairing amplitudes
\cite{roditchev_direct_2015}. Swirling supercurrents and vortices can appear in
structures with strong spin-orbit coupling (SOC), for example when there are
magnetic impurities \cite{pershoguba_currents_2015} or when an in-plane exchange
field is applied \cite{amundsen_supercurrent_2017,bergeret_theory_2020}. In
these cases, spin-triplet pairing amplitudes are created by the
exchange coupling, and modulated by spin-orbit coupling. In analogy to
superfluid He-3 \cite{rantanen_transitions_2023, rantanen_competition_2024,
rantanen_structure_2024}, it might be possible to generate also more exotic
vortex matter in superconducting systems with such triplet pairing amplitudes.

\begin{figure}[t!]
\centering
\includegraphics{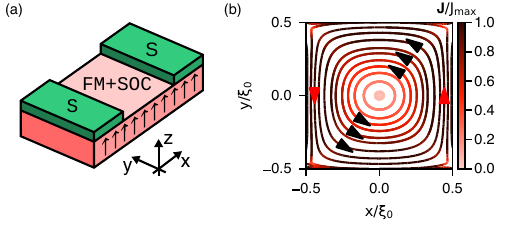}
\caption{\label{fig:setup-summary}
	(a) Sketch of a superconductor-ferromagnetic metal-superconductor device studied here. There is strong spin-orbit coupling in the ferromagnetic layer.
	(b) Circulating current density in the weak link of (a) due to the rotary invariant, computed numerically using the microscopic model. Here no current is applied across the junction.
	Parameters: $(L_x, L_y, J, \alpha) = (\xi_0, \xi_0, \Delta_0, 2.35/\xi_0)$.
}
\end{figure}

Here we introduce the Josephson anomalous vortex (JAV). These appear as a
circulating current density (\autoref{fig:setup-summary}(b)) in two-dimensional
junctions where the Josephson weak link is ferromagnetic and has strong SOC
(\autoref{fig:setup-summary}(a)), as long as no current is applied. JAVs only
appear when the weak link, or gate-defined regions in the weak link, are close
to a coherence length in length and width. The current density is at a maximum
at the edges of the vortex, and vanishes at the center. Correspondingly, at
the center of a vortex the triplet part of the pairing amplitudes vanishes,
leaving only the singlet part; these vortices do not have a core. These
amplitudes are textured in a manner similar to a magnetic vortex.

In this Letter we study a previously unexplored rotary invariant in the free
energy. The rotary invariant is related to a boundary term, and so in the
absence of an applied current generates circulating currents.
Circulating currents due to the rotary invariant arise even when the standard
Lifshitz invariant \cite{agterberg_magnetoelectric_2012} vanishes. We show that
JAVs are not fine-tuned so that they can appear over a range of spin-orbit and
exchange energies. We argue that, due to the magnetic field produced by swirling
currents, JAVs could be detected by scanning magnetometry techniques. We also
suggest that these vortices may be manipulated by Rashba ``islands" within the
weak link via back gates.

Throughout this Letter we use the singlet superconducting gap $\Delta_0$ of the
proximity induced superconductivity in the ferromagnet
as the natural unit of energy, and the superconducting coherence length $\xi_0$
as the unit of length. We restore the physical units when predicting amplitudes
of measurable quantities. In numerical calculations we fix the temperature to $T
= 0.1 \Delta_0/k_{\rm B}$. We use the conventions that Greek subscripts
$\mu,\nu$ run over $(0,1,2,3)$, while Latin subscripts run over $(1,2,3)$.
Repeated indices imply summation.

{\it Magnetoelectric effects without a Lifshitz invariant---} Over the past
decades, it has been shown experimentally and theoretically that SOC allows for
coupling between the magnetic and electric properties of a material
\cite{agterberg_magnetoelectric_2012}. Such coupling is expressed in phenomena
such as the superconducting and Josephson diode effects
\cite{nadeem_superconducting_2023}, helical phases
\cite{hasan_supercurrent_2024}, and the Edelstein and inverse Edelstein effects
\cite{edelstein_magnetoelectric_1995,edelstein_magnetoelectric_2005,dimitrova_theory_2007}.
For weak superconductivity many magnetoelectric effects can be described by the
addition of a Lifshitz invariant in the Ginzburg-Landau free energy
\cite{agterberg_magnetoelectric_2012}. In this Letter we show first from
symmetry that magnetoelectric effects can give rise to  boundary invariants in
addition to the bulk Lifshitz invariant in the most generic free energy. The
most interesting is an invariant that we term the rotary invariant (reported
earlier in Ref. \cite{virtanen_magnetoelectric_2025} but with its significance
not discussed). Such rotary invariants   can appear in a wide range of
superconducting systems when there is no Lifshitz invariant, as long as the
relevant symmetries are satisfied, and generally they describe vortex-like
excitations appearing in inhomogeneous systems or systems with boundaries. With
insight from this general symmetry analysis, we apply a concrete microscopic
model (using quasiclassical methods \cite{virtanen_magnetoelectric_2021})
describing disordered superconductors in the presence of spin-orbit coupling,
derive the microscopic expressions for the invariants, and study the vortices in
parameter regimes much beyond the Ginzburg-Landau theory.

Consider first the superconducting free energy density
\begin{align}
	\mathcal F = \mathcal F_{\rm S} + \mathcal F_{\rm L} + \mathcal F_{\rm R}.
\end{align}
Here $\mathcal F_{\rm S}$ is the usual Ginzburg-Landau free energy \cite{tinkham_introduction_2004},
\begin{align}
	\mathcal F_{\rm S} = a |\psi|^2 + \frac{b}{2} |\psi|^4 + \frac{1}{2}D_{jk} \Pi_j\psi(\Pi_k\psi)^* + \frac{|\vec{B}|^2}{2\mu_0},
\end{align}
with $\psi$ the complex order parameter of the superconductor, $a$ and $b$ the
Ginzburg-Landau constants, $D_{jk}$ the superfluid stiffness symmetric tensor,
and $\vec{B}$ the magnetic field. For compactness we define the momentum $\Pi_j
= -i \hbar \CD_j = -i \hbar \partial_j - 2e A_j$, with $A_j$ the electromagnetic
vector potential. If time-reversal symmetry is broken, and inversion symmetry
intrinsically broken, we can add $d_k$, the Lifshitz invariant vector, via the
second term $\mathcal F_{\rm L}$ \cite{lifshitz_phase_1958}, with
\begin{align}
	\mathcal F_{\rm L} =  d_k [\psi^* \Pi_k \psi + \psi (\Pi_k \psi)^*].
\end{align}
If time-reversal symmetry is broken and the inversion symmetry is broken by
extrinsic inhomogeneity (\eg, via system boundaries, see Appendix A) we can add
the rotary invariant $e_{jk}$, an antisymmetric tensor, via the third term
$\mathcal F_{\rm R}$, with
\begin{align}
	\mathcal F_{\rm R} = i e_{jk} \Pi_j \psi (\Pi_k \psi)^*.
\end{align}
One mechanism for breaking time-reversal and inversion
symmetry comes from the combination
of SOC and exchange field $\vec{h}$, which we consider here.
In the case of
linear in momentum SOC, with $H_{\rm SO} = \mathcal{A}_j^a k_j \sigma^a/(2m)$,
we can define the SU(2) fields $\mathcal{A}_j = \mathcal{A}_j^a \sigma^a$
\cite{gorini2012onsager,bergeret2014spin} and the corresponding field strength
tensor $F_{ij} = -i[\mathcal{A}_i, \mathcal{A}_j]$, valid for position
independent SU(2) fields. Using these definitions, we find that (see Sec.~II in
the supplement \cite{supp}) generally
\begin{equation} d_k = i \kappa \, {\rm Tr}\left\{F_{kj} [\mathcal{A}_j,h_i
\sigma_i]\right\},\quad e_{jk} = \kappa \, {\rm Tr}[F_{jk} h_i \sigma_i],
\end{equation}
where $\kappa=-\frac{7\zeta(3)\nu_{\rm F} \eta}{16\pi^2 k_B^2 T_c^2}$, with
$\nu_{\rm F}$ the density of states per spin projection at the Fermi level,
$\eta = D \ell^2 / (k_{\rm F} \ell)$, $D=v_{\rm F} \ell/2$ the diffusion
constant, $v_{\rm F}$ the Fermi velocity, $k_{\rm F}$ the Fermi wavevector,
$T_c$ is the critical temperature, and $\ell$ the mean free path. $\zeta(n)$ is
the Zeta function. Thus $d_k$ is cubic and $e_{jk}$ quadratic in the strength of
SOC (here we assume weak SOC; in contrast, when $\alpha \, k_{\rm F} \gg
\Delta_0$ the Lifshitz invariant $d_k$ is linear in $\alpha$
\,\cite{hasan_supercurrent_2024}). Moreover, these forms imply that $d_k$ is
non-zero only when the exchange field and $F_{ij}$ have perpendicular
components, whereas $e_{jk}$ requires the presence of their parallel components.
We can hence study the implications of one without the other.

To illustrate this point, we focus on the junction geometry shown in
\autoref{fig:setup-summary}(a). Here we have an exchange field $\vec{h} = J
\hat{\vec{z}}$, and Rashba SOC due to an electric field also in the $z$
direction. The SOC Hamiltonian is $H_{\rm SO} = \alpha (\hat{\vec{z}} \times
\vec{k})\cdot \vec{\sigma}$ and the corresponding field-strength tensor is
$F_{xy} = -F_{yx} = 2\alpha^2 \sigma_z$. Hence, the Lifshitz invariant vanishes:
$d_k=0$, and the rotary invariant is finite: $e_{xy} = -e_{yx}=4 \kappa \alpha^2
J$. Integration by parts of this rotary term results in a boundary term
$\propto{}n_je_{jk}[\psi\partial_k\psi^*-\mathrm{c.c.}]$ that generates
circulating currents perpendicular to the boundary normal~$n_j$.

To investigate the rotary invariant outside of the Ginzburg-Landau regime,
for the rest of this Letter we study a microscopic model for the heterostructure
sketched in \autoref{fig:setup-summary}(a), using the methods described in
Refs.~\onlinecite{virtanen_magnetoelectric_2021,virtanen_nonlinear_2022,kokkeler_universal_2025}
(see Sec.~I in the supplement \cite{supp} for details). From the preceding
analysis we expect a finite rotary invariant (and the related magnetoelectric
effects) for this geometry because the relevant symmetries are satisfied. We
assume that the ferromagnet is weak, so that the quasiclassical approximation is
valid (\ie, the density of states is the same for both up and down spins).
Note that we assume that the superconducting electrodes proximity induce
superconductivity into the ferromagnetic thin film region just below the
electrodes. In general this gap $\Delta_0$ is smaller than the zero-temperature,
zero-field gap of the electrodes. Spin-orbit coupling also increases the
critical field, so we can study the effects of large exchange field, $J \sim
\Delta_0$ \cite{heikkila_thermal_2019}.

Due to the exchange field, the order parameter (\ie, pairing amplitude or
anomalous Green's function) for this system is not scalar, but rather has a spin
structure, $\vec{f} = f_s \sigma_0 + f_i \sigma_i$
\cite{bergeret_long-range_2001}. Here $f_s$ is the singlet part and
$f_i=\vec{f}_t = (f_x, f_y, f_z)$ the triplet part of the anomalous Green's
function. As expected from the Ginzburg-Landau analysis, this microscopic model
also has a rotary invariant. For this model, the rotary invariant involves the
$f_s$ and $f_z$ components of the pairing amplitude (see Sec. I in the
supplement \cite{supp}),
\begin{equation}
	\label{eq:rotary-free-energy}
	\mathcal F_{\rm rot} = i\zeta (\CD f_s \times \CD^* f_z^*) \cdot \hat {\vec{z}}.
\end{equation}
Here $\zeta = \pi \nu_{\rm F} \hbar \eta \alpha^2$; in all numerical results we
assume $\hbar \eta = 0.01 \Delta_0 \xi_0^4$. Note that this term scales as $\sim
1/E_{\rm F}$ and so breaks the quasiclassical symmetry
\cite{silaev_anomalous_2017, virtanen_nonlinear_2022} (that is, the combined
time-reversal and electron-hole symmetries). The
rotary term $\mathcal F_{\rm rot}$ appears as long as there is both SOC and an
exchange field, even when there is no Lifshitz invariant. It generates
circulating current densities,
\begin{align}
	\label{eq:rotary-current}
	\vec{J}_{\rm rot} = 2 e \zeta \left( f_s \curl\, f_z^* + f_s^* \curl\, f_z \right),
\end{align}
with $\curl\, A$ = $(\partial_y A, -\partial_x A)$ and $e$ the elementary
charge. Here $f_s$ varies across the junction due to the proximity effect and
$\vec{f}_t$ is in general finite due to the presence of the exchange field. We
expect the conventional Ginzburg-Landau terms to dominate over the rotary
invariant, so to see the effects of this term there should be no current (\ie,
phase bias $\varphi$) across the junction.

We emphasize that these circulating currents are distinct to those introduced in
Ref.\,\cite{mironov_spontaneous_2017}, which are due to an interfacial SOC
generated surface Lifshitz invariant. While these surface currents can be
similar 
effect here. This depends on the geometry: in \autoref{fig:setup-summary}(a),
the FM/S interface normal is parallel to the exchange field, and the interfacial
Lifshitz invariant vanishes. In contrast, in a lateral junction (where the
ferromagnetic weak link is sandwiched between two superconductors on the left
and right), both the interface and the bulk effects would matter.


\begin{figure}[t!]
\centering
\includegraphics{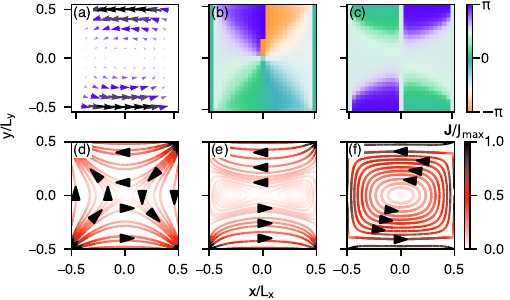}
\caption{\label{fig:antivortices-textures}
	(a) Texture formed from the anomalous Green's functions $f_x, f_y$,
	reminiscent of a magnetic texture.
	(b) Phase winding of the anomalous Green's functions $f_x + i f_y$,
	reminiscent of an Abrikosov vortex.
	(c) Phase texture of the superconducting correlation $f_z$.
	(d) Representative antivortex pattern in the current density.
	(e) Representative mixture between a current antivortex and a vortex patterns.
	(f) Representative vortex.
	Parameters for all panels: $(L_x, L_y, J) = (\xi_0, \xi_0, \Delta_0)$;
	for (a-c): $\alpha = 2/\xi_0$;
	for (d): $\alpha = 1/\xi_0$;
	for (e): $\alpha = 1.55/\xi_0$;
	for (f): $\alpha = 2/\xi_0$;
}
\end{figure}

In \autoref{fig:setup-summary}(b) we plot the current density for a junction of
$L_x \times L_y = (1 \times 1)\xi_0^2$, at zero phase bias. This yields a
representative current vortex, distinguished by the strong current density at
the boundary (as opposed to the center, as in an Abrikosov vortex) and vanishing
current in the center. We compute numerically current densities and pairing
amplitudes for the microscopic model using the code described in
Ref.\,\onlinecite{virtanen_magnetoelectric_2025}. For large junctions we find
some swirling current close to the superconducting terminals (see Sec.~III in
the supplement \cite{supp}), similar to what has been reported in
Ref.\,\onlinecite{bergeret_theory_2020} for in-plane exchange fields. As long as
$\alpha^2 \xi_0^2 \gg \varphi$, the rotary invariant, and thus the circulating
current, dominate over the usual Josephson current. For $\varphi \gtrsim
\alpha^2 \xi_0^2$ the rotary invariant perturbs the Josephson current (see
Sec.~IV in the supplement \cite{supp}).


\begin{figure*}[t!]
\centering
\includegraphics{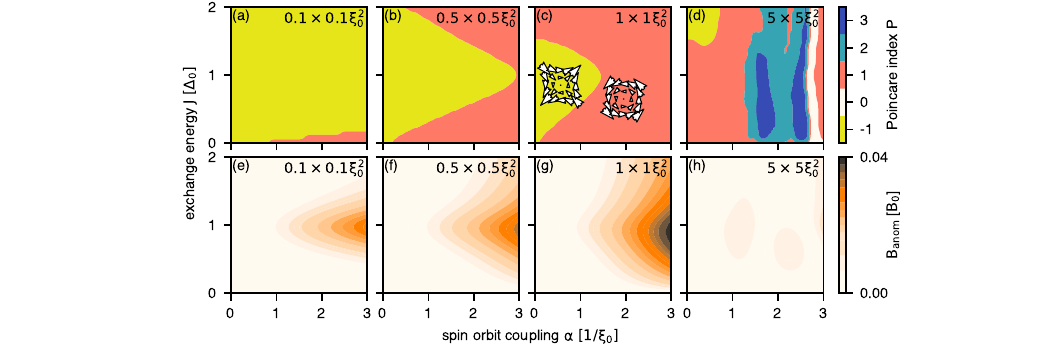}
\caption{\label{fig:phase-diagrams}
		(a-d) Total Poincar\'e index $\mathcal P$ for various junction sizes,
		over a range of exchange energies $J$ and spin-orbit coupling strengths
		$\alpha$. Antivortex patterns in the current correspond to $\mathcal P = -1$
		and vortices to $\mathcal P = +1$, as indicated by the inset sketches.
		Junction size is indicated in the top right hand corner, $L_x \times L_y
		\xi_0^2$.
		(e-h) Magnetic field strength along the $\hat z$ direction at the centre
		of the junction, for the same parameter spaces.
		The Poincar\'e index and magnetic field strength are computed
		from the numerically calculated current densities.
}
\end{figure*}

{\it Antivortex patterns and superconducting textures ---} Often current
vortices in superconductors are accompanied by textures in the phase of the
superconducting order parameter; in this case the anomalous Green's function
(pairing amplitude) $\vec f$. The connection between current vortex patterns and
pairing amplitudes in the case considered here is more complicated. Namely, for
all finite exchange fields $J$ and Rashba spin-orbit coupling strengths $\alpha$
tested we find numerically that the triplet part of the pairing amplitudes is textured. In
\autoref{fig:antivortices-textures}(a) we plot $f_x, f_y$ as a vector field,
forming a swirling pattern reminiscent of a magnetic texture. Equivalently, we
plot in \autoref{fig:antivortices-textures}(b) the complex phase of the $f_x + i
f_y$ anomalous Green's functions, featuring a phase winding reminiscent of
an Abrikosov vortex. Finally, we plot in \autoref{fig:antivortices-textures}(c)
the phase texture of the $f_z$ triplet. This texture does not completely wind,
due to the line defect along $x = 0$. All three textures presented are
representative, with only small variations with $J, \alpha$. In all our
simulations, the amplitudes scale like $|f_s| > |f_z| > |f_x| \sim |f_y|$. From
\autoref{eq:rotary-current} it is clear that the $f_z$ texture generates current
vortex patterns (\autoref{fig:antivortices-textures}(f)). In addition to
vortices we also find antivortex patterns
(\autoref{fig:antivortices-textures}(d)) \cite{footnote_antivortex} in the
current density, and mixtures between the two
(\autoref{fig:antivortices-textures}(e)). As with the vortices, the current
vanishes in the center of an antivortex pattern. The appearance of the latter
patterns is due to the influence of the $f_x, f_y$ textures, and depends subtly
on the small variations of $\vec{f}_t$ with $J,\alpha$.

Because the components of the pairing amplitude respect the 180$\degree$
rotation symmetry of the junction, the triplets $\vec f_t$ change sign across
the origin, and so the triplets (and thus the current) vanishes at the origin.
However, the singlet $f_s$ does not change sign when no current is applied  due
to the symmetric boundary conditions. Thus the singlet is finite at the origin
and these vortices are coreless.

The most natural way to categorize singularities in vector fields (\ie, points
in the weak link where the current density vanishes) is with the
Poincar\'e index. This  is the winding number \cite{guillemin_differential_1974}
\begin{equation}
	\mathcal P = \frac{1}{\pi} \oint_C d \theta,
\end{equation}
with $\theta$ the angle a vector field (such as the current density) makes
relative to the $x$-axis, and $C$ a loop around a singularity. When there are
multiple singularities, the total index is the sum of all $\mathcal P$ for each
singularity. This quantity is a good topological invariant, because small
variations in $\vec J$ do not change $\mathcal P$ \cite{milnor_topology_1997}.
However, because the current density does not vanish far from the
singularity, we see a continuous transition between antivortex and vortex
patterns, rather than a first-order transition.

To gain some intuition for this index, consider the current
densities $\vec J_{\rm V}$ and $\vec J_{\rm AV}$, which approximate the vortex
and antivortex patterns (at least not too close to the boundaries),
\begin{align}
	\label{eq:I-approx}
	&\vec{J}_{\rm V} = J_0 \left( \frac{-y}{\xi_0} \hat{\vec{x}} + \frac{x}{\xi_0} \hat{\vec{y}} \right),
	&\vec{J}_{\rm AV} = J_0  \left(\frac{y}{\xi_0} \hat{\vec{x}} + \frac{x}{\xi_0} \hat{\vec{y}} \right),
\end{align}
Here $J_0 = e \zeta \Delta_0/(\hbar \xi_0)$. By choosing the integration path
$C$ to be a circle, a straightforward calculation shows that $\mathcal P = -1$
for antivortex and $\mathcal P = +1$ for vortex patterns --- this is true
regardless of the vorticity (this can also be seen by noting that the
change in the angle, with fixed integration direction, does not depend on
vorticity).

We plot in \autoref{fig:phase-diagrams}(a-d) the total index over a range of $J,
\alpha$, for 4 different system sizes, $L_x \times L_y = \{0.1 \times 0.1, 0.5
\times 0.5, 1 \times 1, 5 \times 5\} \xi_0^2$. For the systems smaller than $L_x
\times L_y = (5 \times 5) \xi_0^2$ there are only two phases, vortex and
antivortex (as sketched in the inset). For the largest system tested, $(5 \times
5) \xi_0^2$, the phase diagram is richer. For $\alpha > 1.25/\xi_0$ there are
multiple singularities, leading to a total index $\mathcal P_{\rm tot} > 1$.
This results in more complicated patterns in the current density that cannot
be simply categorized as a vortex or an antivortex \cite{supp}. When there is
a single singularity ($\alpha < 1.25/\xi_0$) \autoref{eq:I-approx} is valid
and we find vortex and antivortex patterns in the current density.
The smallest
system tested, $(0.1 \times 0.1) \xi_0^2$, does not in general support
vortices, preferring antivortices.
We find that
vortices are preferred when the system size is close to $L_x = L_y = \xi_0$.
Clearly system size is essential to realizing current vortices, as the
current density is related to gradients of the anomalous Green's
functions $\vec f$. For the same reason the appearance of antivortex rather
than vortex patterns depends subtly on small variations in $\vec f_t$.

Using the Biot-Savart law, we plot in \autoref{fig:phase-diagrams}(e-h) the
magnetic field at the center of the vortex (computed from the numerically
calculated current densities), for the same parameter ranges as (a-d). The
magnetic field scale in the plot is $B_0=\mu_0 \sigma_{\rm D} \Delta_0 t/(4 e
\xi_0) = 1$ mT$\times (t/\xi_0) [\sigma_{\rm D}/(1.6 \cdot 10^7 \text{
	S/m})][\Delta_0/(200 \text{ $\mu$eV})]$, expressed in terms of relatively
typical values for the Drude conductivity $\sigma_{\rm D} = \nu_{\rm F} e^2 D$
in disordered metals, and a typical $\Delta_0$ for Al. We also include the
thickness $t \ll \xi_0$ of the junction. We do not compute this magnetic field
self-consistently (\ie, by feeding back into the electromagnetic vector
potential). As a comparison, we can apply the Biot-Savart law to the
approximation \autoref{eq:I-approx}, yielding the magnetic field at the center
of the vortex
\begin{equation}
	\vec B_{\rm anom} = \mu_0 J_0 \frac{R t}{2 \xi_0},
\end{equation}
with $\mu_0$ the vacuum permeability, and $R$ the radius of the vortex.
Substituting in $R=\xi_0$, we find a maximum magnetic field strength of $B_{\rm
anom} =  2 \pi (R/\xi_0) (\alpha \xi_0)^2 (\eta/\Delta_0 \xi_0^4) B_0$, some 1.5
times larger than what is seen in numerics for vortices in the $(1\times
1)\xi_0^2$ system.

For the $( 1\times 1 )\xi_0^2$ system, which hosts vortices for most of
parameter space, we see the largest $\vec B_{\rm anom}$, increasing with
increasing $\alpha$, and peaking at $J \sim \Delta_0$. This magnetic field is
due to the vortices. For the $( 0.5 \times 0.5 )\xi_0^2$ system we see a weaker
magnetic field, also increasing with $\alpha$ and peaked at $J \sim \Delta_0$.
This magnetic field is due to the mixture of vortex and antivortex patterns
around the phase transition. Surprisingly, for the
$( 0.1 \times 0.1 )\xi_0^2$ system there is also a small magnetic field around
$J \sim \Delta_0$ and $\alpha \sim 3/\xi_0$. While \autoref{eq:I-approx} implies
that for an antivortex pattern there is no circulating current and so no
magnetic field is generated, we find numerically that the antivortices deviate a
little from an ideal antivortex and there is sufficient circulating current to
generate a magnetic field. For the $( 5 \times 5 )\xi_0^2$ system there is
almost no magnetic field. In general, most of the current density is
concentrated close to the superconducting leads, so in the vortex phase $\alpha
< 1.25/\xi_0$ there is very little current and hence only a tiny magnetic field
generated. When there are multiple singularities, $\alpha > 1.25 /\xi_0$, the
currents around each singularity counterpropagate, resulting in the overall
magnetic field vanishing (see Sec. III in the supplement \cite{supp}). For
(e-g), the peak in $B_{\rm anom}$ around $J = \Delta_0$ is due to the fact that
$f_z \sim J / (J^2 + \omega^2)$ (with $\omega$ the Matsubara frequency)
\cite{supp} --- \ie, the magnetic field must peak at some finite $J$ and then
decay for large $J$.


\begin{figure}[t!]
\centering
\includegraphics{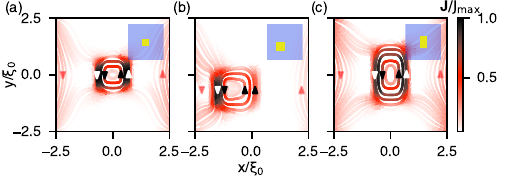}
\caption{\label{fig:manipulation}
	The size (a), position (b), and shape (c) of a JAV could be manipulated by
	defining an island with strong Rashba SOC $\alpha_{\rm island}$ via gates,
	within a weak link with weak SOC $\alpha_{\rm back}$. Inset: sketches of the
	gate defining $\alpha_{\rm island}$ (yellow patch) within the weak link
	(blue square). Parameters: $(L_x, L_y, J, \alpha_{\rm island}, \alpha_{\rm
	back})$ = $(5\xi_0, 5\xi_0, \Delta_0, 2/\xi_0, 1/(2\xi_0))$.
}
\end{figure}

{\it Characterization and manipulation---} Because circulating currents produce
a uniform magnetic field, a natural approach to characterizing these vortices is
via nano-magnetometry. Scanning probe techniques are well-established and can
operate at cryogenic temperatures with high spatial resolution. Magnetic force
microscopy has already been used to study conventional Josephson vortices
\cite{dremov_local_2019, grebenchuk_observation_2020}, and can operate with a
resolution of 10 --- 50 nm \cite{soltys_magnetic-field_2020}, or even atomic
resolution \cite{kaiser_magnetic_2007}. Scanning nitrogen-vacancy center
\cite{welter_scanning_2022} and nanoSQUID methods \cite{vasyukov_scanning_2013}
can now both achieve spatial resolutions of around 50 nm; indeed, recently,
scanning nitrogen-vacancy centers were used to study vortices in NbSe$_2$
\cite{jayaram_probing_2025}. These magnetometry techniques are complicated by
the need to subtract the background ferromagnetic signal. One possible solution
is to include a back gate to control the Rashba interaction
\cite{farzaneh_observing_2024}. One could then effectively switch off SOC to
isolate the background ferromagnetic signal. The absence of SOC can, for
example, be confirmed via the absence of a Josephson diode effect.

Such gate-based techniques for controlling Rashba SOC could also be used to
manipulate these vortices. Consider a large weak link which has only weak (or
absent) Rashba SOC. One can define an island within the weak link with strong
Rashba SOC via a small back gate. We show in \autoref{fig:manipulation} examples
where such Rashba ``islands" can be used to control the size, position, and
shape of a JAV. Such methods could be used to realize JAVs in arbitrarily sized
junctions, overcoming the necessity of defining a square geometry of a coherence
length. Indeed, several different vortices could be defined within a single weak
link using this technique.



{\it Outlook---} We describe the characterization and manipulation of Josephson
anomalous vortices, which appear in Josephson junctions where the weak link is a
weak ferromagnet with strong spin-orbit coupling. The exchange field and
spin-orbit coupling conspire to lead to circulating currents that we associate
phenomenologically to a previously unexplored {\em rotary term} in the free
energy. The same ingredients lead to textures in the triplet components
$\vec{f}_t$.

Josephson anomalous vortices could be found in weak links based on
heterostructures such as ferromagnetic thin film/normal metal bilayers,
ferromagnetic alloys deposited as thin films on metals, or two-dimensional van
der Waals heterostructures, coupled to conventional superconducting electrodes.
Possible thin film bilayers include YIG/InAs or EuS/InAs
\cite{geng_superconductor-ferromagnet_2023}, which must be tuned so that the
magnetization turns out-of-plane (see \eg, Ref. \, \cite{bai_full-scale_2025} and
references therein). Candidate ferromagnetic alloys include Pd/Ni
\cite{kontos_inhomogeneous_2001} or Cu/Ni alloys \cite{ryazanov_coupling_2001}.
The layers must be chosen carefully to minimize the Dzaloshinskii-Moriya
interaction, which will produce a magnetic texture rather than a pure
out-of-plane exchange field. Finally, we suggest that the van der Waals
monolayer magnet CrI$_3$, which has out-of-plane magnetization
\cite{huang_layer-dependent_2017}, could be combined with other van der Waals
materials with strong Rashba spin-orbit coupling, such as MoS$_2$ or WSe$_2$
\cite{shao2016strong}, to find the anomalous vortices.


\begin{acknowledgments}
	The authors acknowledge discussions with G. E. Volovik, E. Strambini, A.
	Dunbrack, and R. Ojajärvi. This work was supported by the European Union’s
	HORIZON-RIA programme (Grant Agreement No. 101135240 JOGATE), managed by the
	Chips Joint Undertaking, and the Research Council of Finland (Contract No.
	355056 and 354735). We acknowledge grants of computer capacity from the
	Finnish Grid and Cloud Infrastructure (persistent identifier
	urn:nbn:fi:research-infras-2016072533).
\end{acknowledgments}


\section{Appendix A: Lifshitz and rotary invariants in Ginzburg-Landau theory}

Here we derive the rotary invariant in the Ginzburg--Landau picture
phenomenologically, and discuss its meaning in more detail.

The expression for the free energy must remain unchanged with respect to various
global transformations that leave the physical system invariant
\cite{lifshitz_phase_1958}. The transformations of the order parameter field are
$\psi\mapsto\psi^*$ (time reversal $\mathcal T$, TR),
$\psi(\vec{r})\mapsto\psi(-\vec{r})$ (inversion $\mathcal I$),
$\psi(\vec{r})\mapsto\psi(R_{ij}r_j)$ (global rotation), and
$\psi(\vec{r})\mapsto{}e^{i\varphi}\psi(\vec{r})$ (global $U(1)$); coefficients
transform keeping $\mathcal{F}$ invariant and real. Further restrictions would
come from crystal symmetry, but we do not consider that here.

Expanding in gradients and expressing the result in terms of real coefficients,
the possible terms with one gradient are
\begin{align}
	\mathcal{F}_1 &= d_k i [\psi \partial_k \psi^* - \mathrm{c.c.}] + c_k \partial_k |\psi|^2
\end{align}
Then, $d_k$ is a TR-odd vector ($\mathcal T, \mathcal I =-,-$) and $c_k$ a
TR-even vector ($\mathcal T, \mathcal I = +,-$). From integration by parts, the
$c_k$ term is equivalent to the Ginzburg--Landau $a|\psi|^2$ term, and a similar
boundary term $n_kc_k|\psi|^2$ where $n_k$ is the boundary normal vector.

The possible terms with two gradients are
\begin{align}
	\nn
	\mathcal{F}_2 =
	&D_{jk} \partial_j \psi \partial_k \psi^*
	+
	i e_{jk} \partial_j \psi \partial_k \psi^*
	+
	f_{jk} \partial_j\partial_k |\psi|^2 \\
	&+
	g_{jk} i \partial_j [\psi \partial_k \psi^* - \mathrm{c.c.}]
\end{align}
where without loss of generality we have taken $D_{jk}=D_{kj}$,
$e_{jk}=-e_{kj}$, and $f_{jk}=f_{kj}$, $g_{jk}=g_{kj}$. Then, $D_{jk}$ is a
TR-even tensor ($\mathcal T, \mathcal I = +,+$) and $e_{jk}$ a TR-odd tensor
($\mathcal T, \mathcal I = -,+$). Similarly: $f_{jk}$ is a TR-even and $g_{jk}$
a TR-odd tensor.

For spatially constant coefficients, the terms aside from the superfluid
stiffness $D_{jk}$ part are total derivatives. From integration by parts, the
$f_{jk}$ term is equivalent to the $c_{k}$ term and a boundary term
$n_jf_{jk}\partial_k|\psi|^2$. The other part $g_{jk}$ is equivalent to a
Lifshitz invariant $(\partial_jg_{jk})i[\psi\partial_k\psi^* - \mathrm{c.c}]$
and a similar boundary term $n_jg_{jk}i[\psi\partial_k\psi^* - \mathrm{c.c.}]$.
The contribution from the $e_{jk}$ term also reduces as
\begin{align}
	\int_M i e_{jk} \partial_j\psi\partial_k\psi^* =
		&- \int_{\partial M} \frac{1}{2} n_j e_{jk} i[\psi\partial_k\psi^* - \psi^*\partial_k\psi] \\
		&+ \int_{M} \frac{1}{2} (\partial_je_{jk}) i[\psi\partial_k\psi^* - \psi^*\partial_k\psi]
	\,.
\end{align}
Consequently, also $e_{jk}$ is equivalent to a bulk Lifshitz invariant
and a similar boundary term.

On a phenomenological level, the $e_{jk}$ term can then always as well be
expressed in terms of a boundary term. However, as shown in Sec.~I of
the supplement \cite{supp}, an equivalent term arises naturally as a bulk contribution in the
microscopic theory, which is valid for arbitrary smooth spatially varying gauge
and exchange fields (\ie, on scales longer than the mean free path). For
interfaces described by such theory, the effective boundary Lifshitz invariant
that is responsible for the vorticity is microscopically generated by the bulk
$e_{jk}$ term.

\section{Appendix B: Forming a vortex due to the rotary invariant in a
superconducting disk}

The rotary invariant can lead to a formation of
superconducting vortices. Consider the effect of a spatially constant rotary
invariant $e_{xy}=-e_{yx}=\epsilon$ in a superconducting disk $M$ of radius $R$,
in the absence of the Lifshitz invariant and for a spatially uniform and
rotationally symmetric superfluid weight $D_{jk}=D_S \delta_{jk}$. The total
free energy  is
\begin{align}
	\nn
	F = &\int_M d^2 r \left(-a |\psi|^2+b |\psi|^4 + D_S |\nabla \psi|^2\right) \\
		&+ \epsilon \int_{\partial M} \hat n \cdot {\rm Im}[\psi^* \nabla \times (\psi \hat u_z)] \\
	\nn
	= &\int_0^{2\pi} d\varphi \int_0^R r dr  \left(-a |\psi|^2+b |\psi|^4 + D_S |\nabla \psi|^2\right) \\
		&+ \epsilon \int_0^{2\pi} d\varphi {\rm Im}[\psi^* \partial_\varphi \psi],
\end{align}
where the second line is written in polar coordinates. Note that we have chosen
a convention where $a>0$ corresponds to the superconducting state.

The free energy of a spatially constant superconducting order parameter
satisfying the Ginzburg-Landau equations, \ie, $\psi=\psi_0 \equiv
\sqrt{a/(2b)}$, is $F_0=-\pi R^2 a^2/(4b)$. Let us check if the rotary invariant
can provide an inhomogeneous state with a lower energy. We make the variational
Ansatz
\begin{equation}
	\psi(r,\varphi) = \psi_0 \tanh(r/\xi) e^{in\varphi},\quad n\in {\mathbb Z},
\end{equation}
ensuring that $\psi$ is uniquely defined at all positions inside the disk. Here
$\xi$ is an Ansatz length scale optimized below. Up to exponentially small
corrections for $R \gg \xi$, the change $\delta F=F-F_0$ reads
\begin{equation}
	\delta F = \frac{\pi a}{b} \left[a \xi^2 c_0  +  D_S  \left(c_1+n^2 \log \frac{c_2 R}{\xi}\right) + \epsilon n\right],
\end{equation}
where $c_0=(1+20 \log 2)/12$, $c_1 =(4 \log 2-1)/3$ and $c_2\approx 1.25$. The
term proportional to $a$ comes from the combination of the $a$ and $b$ terms in
the free energy density, when using the particular value for $\psi_0$.
Minimizing this with respect to $\xi$ yields $\xi = |n|\sqrt{D_S/(c_0 a)}$ and
for this minimum scale the free-energy change reads
\begin{equation}
	\delta F = \frac{\pi a}{b} \left\{D_S[c_1+n^2 (1+\log \frac{c_2 R}{\xi})]+\epsilon n\right\}.
\end{equation}

This variational Ansatz shows that the rotary invariant can drive the system to
an inhomogeneous vortex state. The transition to the vortex state (\ie, with
$\delta F <0$) for $\epsilon>0$ takes place with $n=-1$, when
\begin{equation}
	\label{eq:vortextransitionlimit}
	\epsilon \ge D_S\left(1+c_1 + \log \frac{c_2 R}{\xi}\right).
\end{equation}
In other words, the vortex transition requires that the rotary invariant
essentially exceeds the superfluid weight.

Note that this result is based on a crude variational Ansatz, so
\autoref{eq:vortextransitionlimit} is an approximate estimate for the critical
value of the rotary invariant.

\bibliography{josephson-anomalous-vortices}

\begin{thebibliography}{46}
\bibitem [{foo()}]{footnote_antivortex}%
  \BibitemOpen
  \href@noop {} {}\bibinfo {note} {``Antivortex pattern" refers here to the
  behavior of the charge current vector field, with a meaning similar to that
  used for magnetic textures. The meaning differs from antivortices in a scalar
  field, e.g., when discussing circulation of Abrikosov vortices, as it refers
  to the sign of the vector field Poincare index, $\mathcal P = +1$ for vortex
  and $\mathcal P = -1$ for antivortex, instead of the sign of the winding or
  circulation.}\BibitemShut {Stop}%
\end{thebibliography}

\end{document}